\documentclass[10pt]{amsart}

\title[Line Bundle Cohomology and Applications to String Phenomenology]{Recent Developments in Line Bundle Cohomology and Applications to String Phenomenology}

\author[Brodie]{Callum Brodie}
\address{Department of Physics, Robeson Hall, Virginia Tech, Blacksburg, VA 24061, USA}
\email{callumb@vt.edu}

\author[Constantin]{Andrei Constantin}
\address{Rudolf Peierls Centre for Theoretical Physics, University of Oxford, Parks Road, Oxford, OX1 3PU, UK}
\email{andrei.constantin@physics.ox.ac.uk}

\author[Gray]{James Gray}
\address{Department of Physics, Robeson Hall, Virginia Tech, Blacksburg, VA 24061, USA}
\email{jamesgray@vt.edu}

\author[Lukas]{Andre Lukas}
\address{Rudolf Peierls Centre for Theoretical Physics, University of Oxford, Parks Road, Oxford, OX1 3PU, UK}
\email{andre.lukas@physics.ox.ac.uk}

\author[Ruehle]{Fabian Ruehle}
\address{Department of Physics \& Department of Mathematics, Northeastern University, 360 Huntington Avenue, Boston MA 02115, USA}
\address{The NSF AI Institute for Artificial Intelligence and Fundamental Interactions}
\email{f.ruehle@northeastern.edu}

\usepackage[citation-order]{amsrefs}
\usepackage{graphicx}
\usepackage{color}
\usepackage{amsfonts}
\usepackage{amssymb}
\usepackage{amscd}
\usepackage{array}
\usepackage{subfigure}
\usepackage{xypic}
\usepackage[margin=2.5cm]{geometry}
\usepackage{booktabs}
\usepackage{hyperref}
\usepackage{longtable}
\usepackage{pdflscape}
\usepackage{colortbl}

\allowdisplaybreaks[3]
\newcommand{\cicy}[2]{\begin{matrix} #1\end{matrix}\!\left[\begin{matrix}#2 \end{matrix}\right]}
\newcommand{\IP}{\mathbb{P}}

\newcommand{\cO}{\mathcal{O}}

\theoremstyle{plain}

\newtheorem*{lem*}{Lemma}

\theoremstyle{definition}

\newtheorem*{remark*}{Remark}

\newcommand{\floor}[1]{\left\lfloor{#1}\right\rfloor}
\newcommand{\ceil}[1]{\left\lceil{#1}\right\rceil}

\newcommand{\mc}{\mathcal}

\newcommand{\divcls}[1]{\left[{#1}\right]}			

\newcommand{\cK}{{\mathcal K}}
\newcommand{\cM}{{\mathcal M}}

\begin{document}

\begin{abstract}
Vector bundle cohomology represents a key ingredient for string phenomenology, being associated with the massless spectrum arising in string compactifications on smooth compact manifolds. Although standard algorithmic techniques exist for performing cohomology calculations, they are laborious and ill-suited for scanning over large sets of string compactifications to find those most relevant to particle physics. In this article (based on the second author's lecture at the Nankai Symposium, August 2021) we review some recent progress in deriving closed-form expressions for line bundle cohomology and discuss some applications to string phenomenology. 
\end{abstract}

\maketitle

\section{Introduction}

Computing cohomology is a crucial and time-consuming step in the derivation of the spectrum of low-energy particles resulting from string compactifications. The state-of-the-art consists of computer implementations of algorithmic methods, based on \v{C}ech cohomology and spectral sequences  \cite{cicypackage, cohomCalg:Implementation, cicytoolkit}. However, these methods are computationally intensive and provide little insight into the origin of the results. 
Recently, it has been shown that in many cases of interest in string theory topological formulae for line bundle cohomology exist, providing mathematical shortcuts that can reduce the time needed for deciding the physical viability of a string compactification from several months of computer algebra to the split of a second. 

Line bundles feature in many string theory contexts and have proven to be a fruitful setting for realistic phenomenology \cite{Distler:1987ee, Blumenhagen:2006ux, Blumenhagen:2006wj, Anderson:2011ns, Anderson:2012yf, Anderson:2013xka, Buchbinder:2014sya, Buchbinder:2014qda}. Moreover, they represent building blocks for higher-rank vector bundles, such as monad or extension bundles \cite{Kachru:1995em, Blumenhagen:2005ga, Anderson:2008ex, Anderson:2008uw, Anderson:2009mh, He:2009wi}. Topological formulae for line bundle cohomology were initially discovered empirically, through a combination of direct observation \cite{Constantin:2018otr, Buchbinder:2013dna, Constantin:2018hvl, Larfors:2019sie, Brodie:2019pnz} and machine learning techniques \cite{He:2017aed,Ruehle:2017mzq,Klaewer:2018sfl, Brodie:2019dfx} and covered examples from several classes of complex manifolds of  dimensions two and three. The main observation was that the Picard group decomposes into disjoint regions, called {\itshape cohomology chambers}, in each of which the cohomology function is polynomial or very close to polynomial, a pattern observed both for the zeroth and the higher cohomologies, with a different chamber structure emerging for each type of cohomology. The number of regions often increases dramatically with the Picard number. 

It is well-known that given a complex manifold $X$ and a holomorphic bundle $V$, the Euler characteristic $\chi(X,V)$, that is the alternating sum of cohomologies $h^i(X,V)$ can be equated to a purely topological index:  
\vspace{-2pt}
\begin{equation*}
\chi(X,V) = \sum_{i=0}^{{\rm dim}(X)} (-1)^ih^i(X,V) =  \int_X {\rm ch}(V) \cdot {\rm td}(X)~,
\vspace{-2pt}
\end{equation*}
which is the statement of the Hirzebruch-Riemann-Roch theorem. 
Similar formulae have now been found to hold for each individual cohomology dimension $h^i(X,V)$, for large classes of complex manifolds and abelian vector bundles. 

\section{Dimension Two}

The first advancement in understanding the mathematical origin of these formulae came through the study of line bundles on several classes of complex surfaces widely used in string theory, including compact toric surfaces, weak Fano surfaces (generalised del Pezzo surfaces), and K3 surfaces~\cite{Brodie:2019ozt, Brodie:2020wkd}. Note that for surfaces it suffices to study the zeroth cohomology function $h^0(X,V)$, from which the formulae for the first and second cohomologies then follow by Serre duality and the Riemann-Roch theorem. 

Line bundle cohomology formulae on these families of surfaces arise through a combination of Zariski decomposition and vanishing theorems. Zariski decomposition is the statement that on any complex projective surface $X$, any effective divisor $D$ can be uniquely decomposed as $D=P+N$, where $P$ is nef, $N$ is effective, and $P$ intersects no components in the curve decomposition of $N$. In general, the divisors $P$ and $N$ are rational. When $D$ is integral, its class defines an effective line bundle and the following equality holds
\begin{equation}\label{eq:dP2}
h^0\big(X, \mc{O}_X(D)\big) = h^0\big(X, \mc{O}_X(\floor{P})\big) = h^0\big(X, \mc{O}_X(\ceil{P})\big) \,,
\end{equation}
where $\floor{P}$ and $\ceil{P}$ are the maximal integral subdivisor of $P$ and, respectively, the minimal integral divisor that has $P$ as a subdivisor. The classes $\divcls{P}$, $\divcls{\floor{P}}$ and $\divcls{\ceil{P}}$ depend only on the class $\divcls{D}$, and, crucially, can be computed purely from the intersection properties of $D$. In particular this computation requires knowledge of the Mori cone and the intersection form on $X$.
If the cohomology on the right-hand side of Equation~\eqref{eq:dP2} can be computed more easily than on the left, the relation becomes practically useful. 

\subsection{Zariski chambers}
The map $[D]\rightarrow [P]$ gives rise to a {\itshape Zariski chamber decomposition} of the effective cone~\cite{Bauer04}. When $D$ is nef it follows that $N$ is trivial and $D=P$, hence the nef cone is a chamber in itself. Furthermore, every face $F$ of the nef cone not contained in the boundary of the Mori cone gives rise to a Zariski chamber $\Sigma_F$, by translating the face $F$ along the dual Mori cone generators. Thus, if $D$ is an effective divisor within the chamber $\Sigma_{i_1\ldots i_n}$ obtained by translating the codimension $n$ face $F$ of the nef cone along the set of Mori cone generators $R = \{\cM_{i_1},\cM_{i_2},\ldots, \cM_{i_n}\}$ orthogonal to $F$ with respect to the intersection form, then $P$ lies on $F$ and
\begin{equation}
P = D - \sum_{k=1}^n\, (-D\cdot \cM_{i_k , R}^\vee) \,\cM_{i_k}  \,,
\end{equation}
where the dual $\cM_{i_k,R}^\vee$ is an effective divisor with support $R$ defined by $ \cM_{i_k,R}^\vee\cdot  \cM_{i_m} = -\delta_{km}$, $\forall \cM_{i_m} \in R$. 

\subsection{Cohomology chambers}
While Zariski chambers exist for every smooth complex projective surface, whether these become cohomology chambers or not depends on the presence of an appropriate vanishing theorem that interacts well with the floor/ceiling operations in Eq.~\eqref{eq:dP2}. If the image of a Zariski chamber under the map $\divcls{D} \to \divcls{\floor{P}}$ or under the map $\divcls{D} \to \divcls{\ceil{P}}$ is covered by a vanishing theorem, the index function can be `pulled back' to give a single function for zeroth cohomology throughout the Zariski chamber, which is now also a {\itshape cohomology chamber}. 
This is indeed the case for the following classes of surfaces:
\begin{itemize}
\item[a.] Generalised del Pezzo surfaces (including del Pezzo surfaces). The Kawamata-Viehweg vanishing theorem is relevant in this case and, in the above notation
\begin{equation}
h^0\big(X, \mc{O}_X(D)\big) = \chi\bigg(X, \mc{O}_X\Big(D - \sum_{k=1}^n\, \ceil{-D\cdot \cM_{i_k,R}^\vee} \,\cM_{i_k}\Big)\bigg) \,.
\label{eq:gen_pullb1}
\end{equation}
\item[b.] Projective toric surfaces. Demazure's vanishing theorem is relevant in this case and 
\begin{equation}
h^0\big(X, \mc{O}_X(D)\big) = \chi\bigg(X, \mc{O}_X\Big(D - \sum_{k=1}^n\, \floor{-D\cdot \cM_{i_k,R}^\vee} \,\cM_{i_k}\Big)\bigg) \,.
\label{eq:gen_pullb2}
\end{equation}
\item[c.] K3 surfaces. The Kawamata-Viehweg vanishing theorem can be used in this case too, however, this gives cohomology formulae only in the interior of the effective cone, leading to the same equality as in Eq.~\eqref{eq:gen_pullb1}. On the boundary of the effective cone, for the integral divisors $D$ whose support has negative definite intersection matrix it follows that the positive part is trivial $P=0$ so that $h^0\big(X,\mc{O}_X(D)\big) = h^0\big(X,\mc{O}_X\big) = 1$. In general this determines the cohomology on some but not all the faces of the Mori cone. The cohomologies of the remaining line bundles on the boundary require a separate treatment. 
\end{itemize}

\subsection{Example}
\begin{figure}[h]
  \begin{center}
  \includegraphics[scale=0.34]{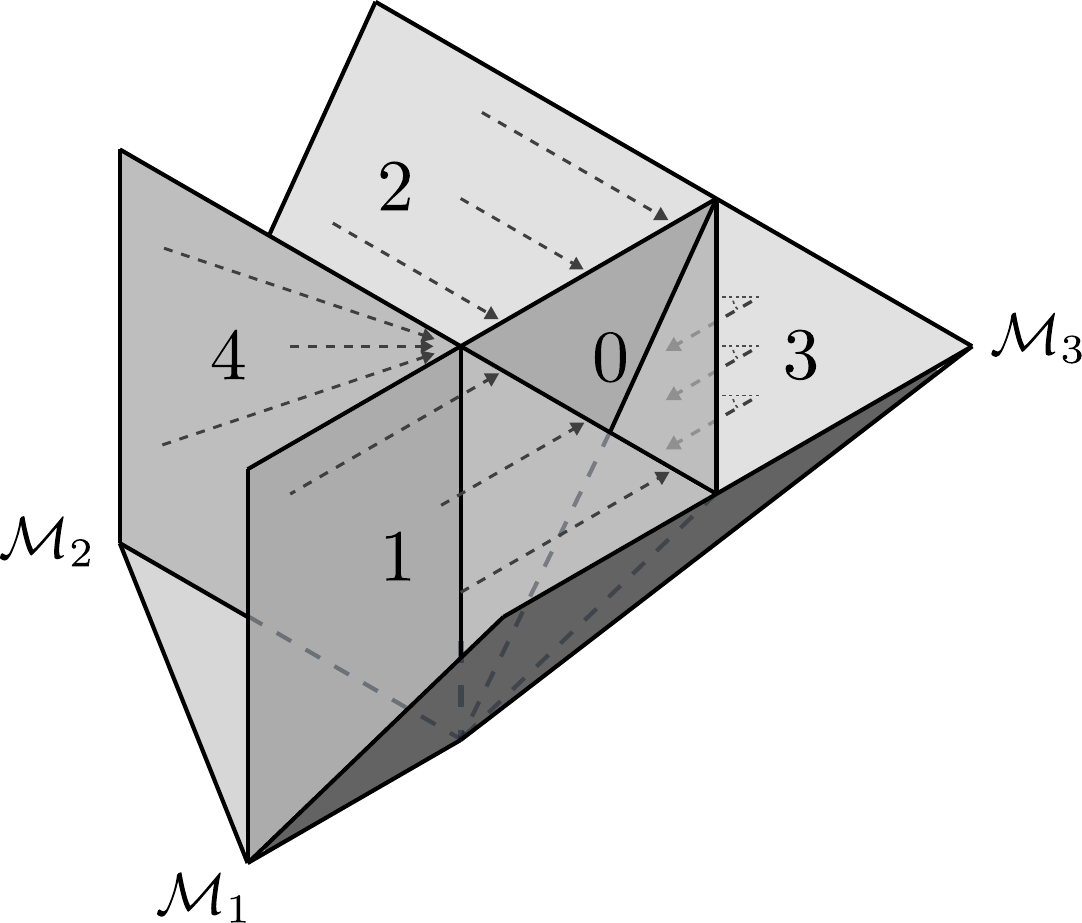}
  \end{center}
\caption{\itshape\small Zeroth cohomology chamber structure of the effective cone of dP$_2$. The projection arrows correspond to the map $\divcls{D} \to \divcls{\ceil{P}}$. The nef cone is the chamber labelled as $0$.}
  \label{dp2_piclat_numb_shifts}
\end{figure}

We illustrate the above discussion with a simple example of a del Pezzo surface of degree~$7$, obtained by blowing up $\IP^2$ at two generic points and denoted ${\rm dP}_2$ in the Physics literature. The Picard lattice of dP$_2$ is spanned by the hyperplane class $H$ of $\IP^2$ and the two exceptional divisor classes $E_1$ and $E_2$ resulting from the two blow-ups. The effective cone (Mori cone) is generated by $\cM_1 = E_1$, $\cM_2 = E_2$, and $\cM_3 = H-E_1-E_2$. All three generators are rigid, satisfying $\cM_i^2=-1$.

The Zariski/cohomology chamber structure of the effective cone is shown in Figure~\ref{dp2_piclat_numb_shifts}. The nef cone has three co-dimension $1$ faces and one co-dimension $2$ face not shared with the Mori cone. Each of these gives rise to a cohomology chamber, four in total. The zeroth cohomology is given by the locally polynomial function
\begin{equation*}
h^0\left(\mathrm{dP}_2,\mc{O}_{\mathrm{dP}_2}(D)\right) = \chi\Big(\mathrm{dP}_2, \mc{O}_{\mathrm{dP}_2}\big(D - \sum_{i=1}^3 \theta( - D \cdot \cM_i ) \, (-D \cdot \cM_i)\cM_i\big) \Big) \,,
\label{eq:dp2_h0_alt}
\end{equation*}
where $\theta(x)$ equals one for $x \geq 0$ and zero otherwise.

\section{Dimension Three}
Calabi-Yau three-folds provide qualitatively new phenomena for line bundle cohomology. Several examples of low Picard number Calabi-Yau threefolds realised as complete intersections in products of projective spaces (CICY threefolds) or hypersurfaces in toric varieties were studied in Refs.~\cite{Brodie:2020fiq, Brodie:2021ain, Brodie:2021nit}. 
As in the case of complex surfaces, in these examples the zeroth cohomology determines a decomposition of the effective cone into polyhedral chambers, in each of which the zeroth cohomology can be expressed as a topological index. The chambers were understood to be either:
\begin{itemize}
\item[(i)] Images of K\"ahler cones of birational models of $X$ under the natural identifications of the Picard groups. Inside such chambers and along their common walls the zeroth cohomology can be equated to the Euler characteristic on the corresponding flopped manifold, due to Kodaira's vanishing theorem, the Kawamata-Viehweg vanishing theorem and the invariance of the number of global sections of a line bundle under a flop. The union of these (images of) K\"ahler cones is known in the Physics literature as the {\itshape extended K\"ahler cone} and in the Mathematics literature as the {\itshape movable cone}.
\item[(ii)] Zariski chambers where a cohomology-preserving projection operates in a similar way as in the two-dimensional case. The existence of a Zariski chamber neighbouring a boundary of the (extended) K\"ahler cone is signalled by the shrinking of an effective divisor to zero volume as computed with respect to the K\"ahler form on the corresponding flopped manifold. 
\end{itemize}

The appearance of the first type of chambers is novel and is related to the fact that in dimension 3 minimal models are not unique. Calabi-Yau threefolds have trivial and hence nef canonical bundle, therefore they are minimal models. It is well-known that in  dimension 3 minimal models are not unique in general, but any two distinct minimal models are birationally related by a sequence of flops, being isomorphic outside subsets of codimension~2.

The zeroth line bundle cohomology encodes the information about the flops connecting the birational models of the manifold, as well as about the Gromov-Witten invariants counting the  number of flopping curves. 
We discovered that the vast majority of known Calabi-Yau threefolds admit flops, while many of these flop to manifolds isomorphic to themselves \cite{Brodie:2021toe}. In particular, many threefolds admit infinite sequences of flops and have an infinite number of contractible rational curves.
\begin{center}
\begin{figure}[h]
\begin{center}
\includegraphics[width=5.8cm]{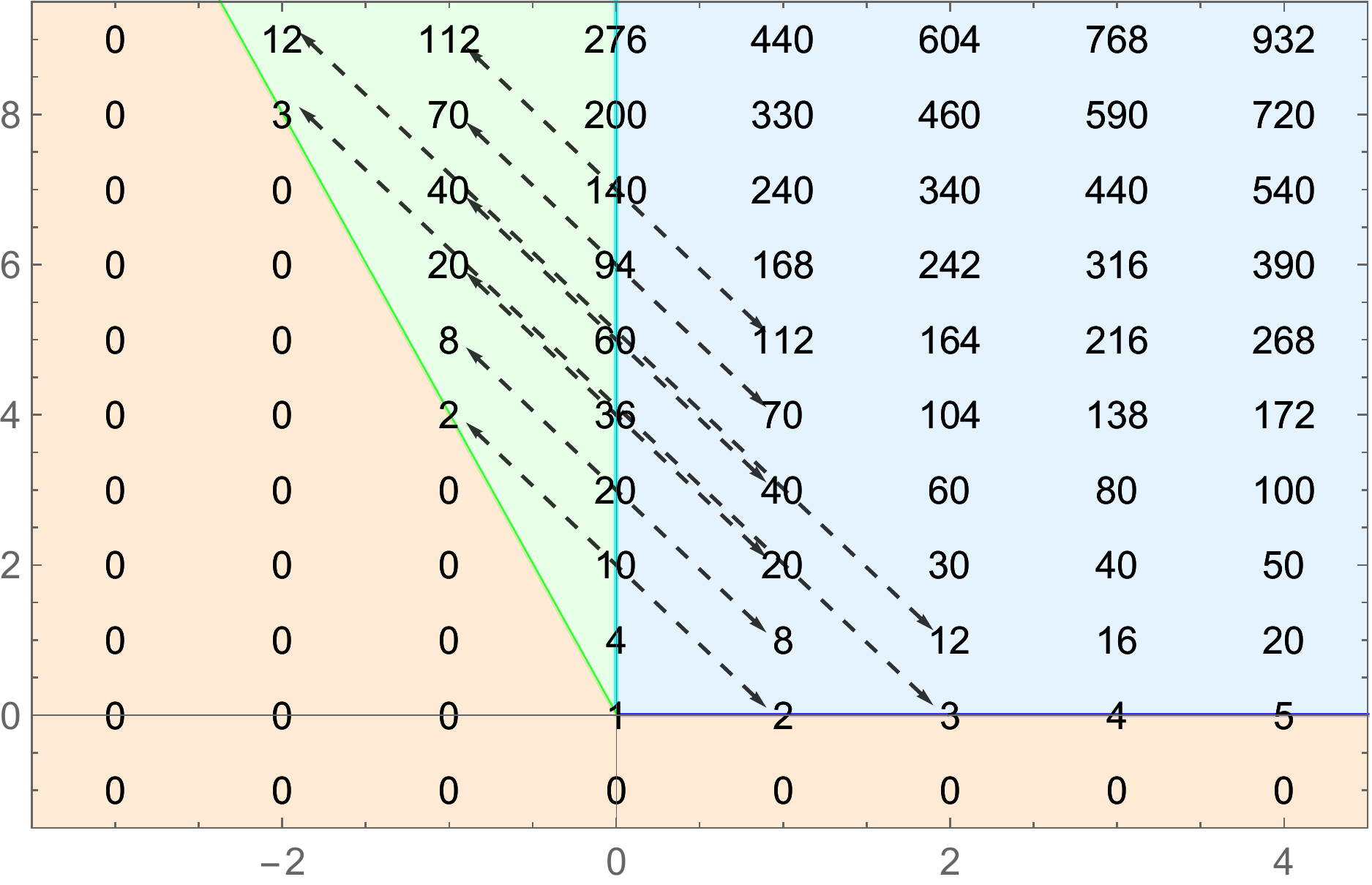}
\hspace{12pt}
\includegraphics[width=5.8cm]{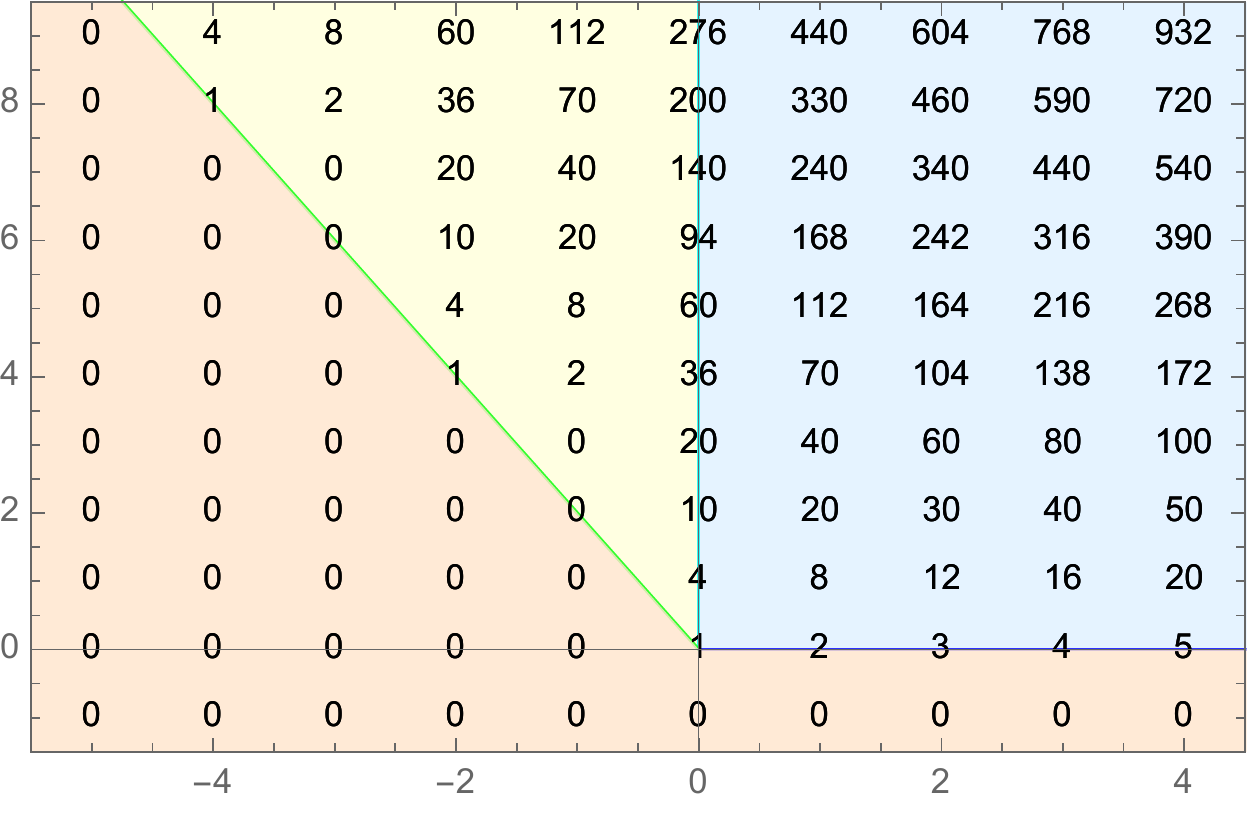}
\vspace{-7pt}
\caption{\itshape Zeroth line bundle cohomology for generic (left) and tuned (right) Calabi-Yau hypersurface  of bi-degree $(2,4)$ in $\mathbb{P}^1\times\mathbb{P}^3$.}
\label{fig:X7887generic}
\vspace{-12pt}
\end{center}
\end{figure}
\end{center}

\subsection{Example}
For illustration, consider a  Calabi-Yau threefold $X$ with two K\"ahler parameters, defined as the zero locus of a generic  homogeneous polynomial of bi-degree $(2,4)$ in $\mathbb{P}^1\times\mathbb{P}^3$. This is the manifold with identifier $7887$ in the CICY list~\cite{Candelas:1987kf,Green:1987cr} with configuration matrix 
\begin{equation}\label{conf7887}
\cicy{\IP^1 \\ \IP^3}{\,2~\, \\ \,4~\,}^{2,86}
\end{equation}
where the superscripts indicate the Hodge numbers $(h^{1,1}(X),h^{2,1}(X))$. The manifold has been recently studied in Ref.~\cite{Brodie:2020fiq} and earlier on in Ref.~\cite{Ottem:2015}. 
The structure of the zeroth line bundle cohomology is shown in the left plot of Figure~\ref{fig:X7887generic}. The blue cone corresponds to the K\"ahler cone of $X$ and here the zeroth cohomology is given by $h^0(X,L)=\chi(X,L)$ due to Kodaira's vanishing theorem. The green cone corresponds to the K\"ahler cone of the flopped manifold~$X'$ and here $h^0(X,L)=h^0(X',L')=\chi(X',L')$ where $L'$ is the line bundle on $X'$ corresponding to $L$ on~$X$. The wall separating the two K\"ahler cones, excluding the origin, is covered by the Kawamata-Viehweg vanishing theorem and here $h^0(X,L)=\chi(X,L)=\chi(X',L')$. 
The line bundles lying along the two boundaries of the effective cone require a separate discussion. Thus, if $L = \cO_X(k_1[D_1]+k_2[D_2])$ is a line bundle with first Chern class written in the basis $\{[D_1],[D_2]\}$ consisting of the restrictions to $X$ of the hyperplane classes on $\IP^1$ and $\IP^3$ (which are simultaneously generators of the Picard group and of the K\"ahler cone of $X$), then $h^0(X,L) = \chi(\IP^1,\cO_{\IP^1}(k_1))$ if $k_1>0,~k_2=0$ and  $h^0(X,L) =\chi(\IP^1,\cO_{\IP^1}(-k_1))$, if $k_1<0$ and $k_2=-4k_1~$.
 Outside of these regions the zeroth cohomology is trivial. The cohomology formulae are summarised below:
\begin{equation*}
\begin{tabular}{ l | c}
 {\rm region in eff.~cone}		&~$ h^0(X,L=\cO_X(k_1[D_1]+k_2[D_2]))$ \\[0pt]
\hline
$\cK(X)$ &~ $\chi(X,L)$ \\[0pt]
$\cK(X')$ &~ $\chi(X',L')$ \\[0pt]
$k_1=0,~k_2>0$ &~ $\chi(X,L)=\chi(X',L')$ \\[0pt]
$k_1<0,~k_2=-4k_1~$ & $\chi(\IP^1,\cO_{\IP^1}(-k_1))$\\[0pt]
$k_1>0,~k_2=0$ & $\chi(\IP^1,\cO_{\IP^1}(k_1))$\\[0pt]
$k_1=k_2=0$ & $1$
\end{tabular}
\label{eq:7887_formulae}
\end{equation*}
 
The flop $X\rightarrow X'$ can be constructed explicitly as follows. Let $[x_0:x_1]$ denote the homogeneous coordinates on the $\IP^1$ and $[y_0:y_1:y_2:y_3]$ the coordinates on the $\IP^3$. The hypersurface equation reads
\begin{equation}\label{eq:7887}
\begin{aligned}
x_0^2 P(y) &+ x_0x_1 Q(y) +x_1^2 R(y) =0~,
\end{aligned}
\end{equation}
where $P,Q$ and $R$ are generic quartic polynomials in the $y$-coordinates. For a generic $y\in\IP^3$ the equation gives two solutions for $x$. On the other hand, when $P(y)=Q(y)=R(y)=0$, which happens at $64$ points in $\IP^3$, the equation admits an entire $\IP^1$ as a solution. By contracting these curves one obtains a singular threefold. The flopped threefold $X'$ is obtained by resolving these singularities into new rational curves. Explicitly, $X'$ is given by the hypersurface in $\IP^1\times \IP^3$
\begin{equation}\label{eq}
\begin{aligned}
x_0^2 R(y) &+ x_0x_1 Q(y) +x_1^2 P(y) =0~,
\end{aligned}
\end{equation}
which is just Eq.~\eqref{eq:7887} with the roles of $P$ and $R$ swapped. 
The  manifold $X'$ is isomorphic to the original one, as reflected by the $\mathbb Z_2$-symmetry of the cohomology data in the left plot of Figure~\ref{fig:X7887generic}. The threefolds $X$ and $X'$ share not only the same Hodge numbers, but also the same triple intersection numbers and $c_2$-form when written in the basis of generators of their respective K\"ahler cones.

\subsection{Complex structure dependence}\label{sec:CS}
An important aspect of the cohomology formulae is their complex structure dependence. In the previous example, when the quartic polynomials $P, Q$ and $R$ in Eq.~\eqref{eq:7887} are generic, the cohomology data corresponds to the situation described above. On the other hand, when all the coefficients in $Q$ vanish there is a dramatic change in the structure of the zeroth line bundle cohomology, as shown in the plot on the right in Figure~\ref{fig:X7887generic}. 
The defining equation is now
\begin{equation}\label{eq:7887tuned1}
\begin{aligned}
x_0^2 P(y) &+ x_1^2 R(y) =0~,
\end{aligned}
\end{equation}
where $P$ and $R$ are generic. At a generic point $y\in\IP^3$, the equation gives two solutions for $x$. However, when $P(y)=R(y)=0$, which corresponds to a curve in $\IP^3$, the solution is an entire $\IP^1$. The $\IP^1$-fibration over this curve gives rise to a rigid divisor $\Gamma$ in the class $-2[D_1]+4[D_2]$. Note that $h^0(X,\cO_X(\Gamma))=1$ and the volume of $\Gamma$ shrinks to zero when the K\"ahler form approaches the relevant boundary of $\cK(X)$. Thus at this locus in complex structure moduli space there are no isolated rational curves that can be flopped.

Zeroth cohomology formulae can be obtained as follows. As before, inside $\cK(X)$ and along the boundary separating $\cK(X)$ from $\Sigma$, excluding the origin, we have $h^0(X,L)=\chi(X,L)$, which follows from Kodaira vanishing inside $\cK(X)$, and Kawamata-Viehweg vanishing for the big line bundles on the boundary. 

The presence of the rigid divisor $\Gamma$ gives rise to a Zariski chamber.
$\Gamma$ is part of the fixed locus of every linear system in the cone $\Sigma$, as indicated by the fact that the cohomology data inside $\Sigma$ is invariant under $\Gamma$-shifts. The presence of $\Gamma$ and the amount by which it is present can be detected by intersection properties, as follows. 
Let $[C_1], [C_2]$ denote the curve classes dual to $[D_1], [D_2]$, such that $[C_i]\cdot [D_j] = \delta_{ij}$. Let $D$ be an effective divisor on $X$ and assume it has a Zariski decomposition $D = P + N$, where $P$ and $N$ are $\mathbb Q$-divisors, $N$ is effective and $P$ is nef, that is $[P]\in \overline\cK(X)$. If $D$ itself is nef, the Zariski decomposition is trivial. Thus we assume $[D]\in\Sigma$. In this case, $[P]$ lies on the boundary shared by $\Sigma$ and $\overline\cK(X)$. From general properties of Zariski decomposition we know that
\begin{equation}
h^0(X,\cO_{X}([D])) = h^0(X, \cO_{X}([\floor{P}]))~,
\end{equation}
where $\floor{P}$ is the integral divisor obtained by rounding down all the coefficients in the expansion of $P$. In terms of divisor classes, the round down class $[\floor{P}]$ is well-defined when $[D]$ is an integral class. The divisor~$N$ is a rational multiple $N = \gamma \Gamma$ of the rigid divisor $\Gamma$ and $[P]$ is a rational multiple of $[D_2]$. Since $D= P+N$ and $D_2\cdot C_1=0$, it follows that $D\cdot C_1 = N\cdot C_1= \gamma\, \Gamma \cdot C_1$, from which $\gamma = D\cdot C_1/ \Gamma \cdot C_1$. Consequently, 
\begin{equation}
\begin{aligned}
h^0(X,~&\cO_{X}([D])) = h^0(X, \cO_{X}([D]{-}\ceil{\gamma}[\Gamma]))~ = \chi\left(X, \cO_{X}\left([D]-\ceil{\frac{D\cdot C_1}{\Gamma\cdot C_1}}[\Gamma]\right)\right).
\end{aligned}
\end{equation}
With $[D]\ = k_1[D_1]+k_2 [D_2]$ it is straightforward to compute $\gamma = -k_1/2$. The zeroth line bundle cohomology formulae are summarised below: 
\begin{equation*}
\begin{tabular}{ l | c}
 {\rm region in eff.~cone}		&~$ h^0(X,L=\cO_X([D]=k_1[D_1]+k_2[D_2]))$ \\[4pt]
\hline
$\cK(X)\cup \left( k_1=0, k_2>0\right)$ &~ $\chi(X,L)$ \\[4pt]
$\overline\Sigma$ & ~ $\chi\left(X, \cO_{X}\left([D]-\ceil{\frac{D\cdot C_1}{\Gamma\cdot C_1}}[\Gamma]\right)\right)$ \\[4pt]
$k_1>0,~k_2=0$ & $\chi(\IP^1,\cO_{\IP^1}(k_1))$\\
$k_1=k_2=0$ & $1$
\end{tabular}
\label{eq:7885_formulae}
\end{equation*}

\begin{figure}[h]
\begin{center}
\includegraphics[width=2.4cm]{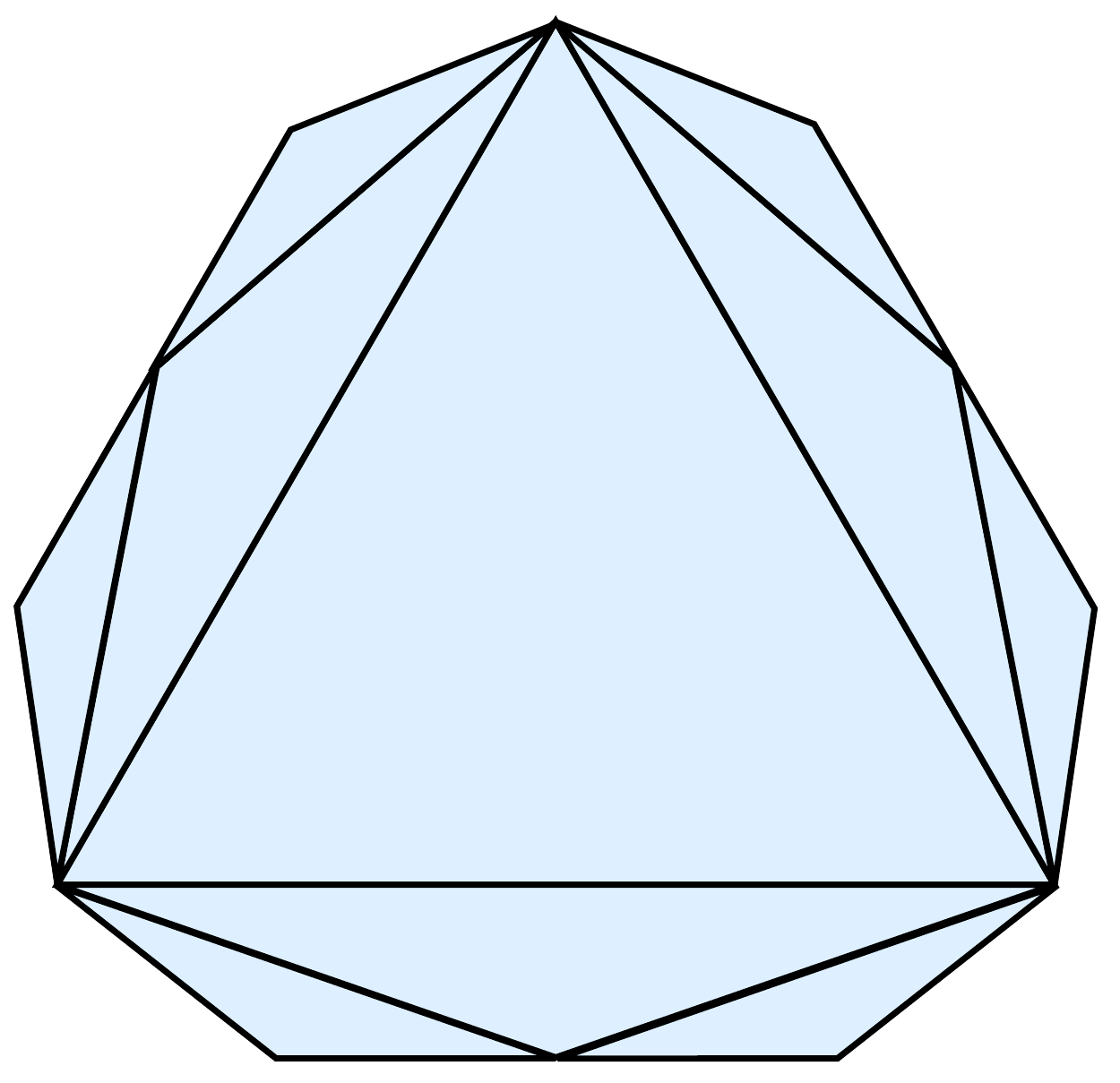}
\vspace{-7pt}
\caption{\itshape Schematic representation of the extended K\"ahler cone of the CICY manifold 7447. There are infinitely many subcones, though only the largest ones are included here.}
\label{KX_7447}
\end{center}
\end{figure}

\subsection{Infinite flop chains.} Infinite flop chains arise whenever isomorphic flops exist through at least two different boundaries of the K\"ahler cone. In this case, the extended K\"ahler cone consists of an infinite number of sub-cones, 
and when $h^{1,1}(X)>2$ it has infinitely many edges (provided that the birational automorphism group is infinite). When $h^{1,1}(X)=2$ the envelope of the extended K\"ahler cone is an irrational two-dimensional cone. 
In Refs.~\cite{Brodie:2021ain, Brodie:2021nit} infinite flop chains were studied in relation to the swampland distance conjecture and the Kawamata-Morrison conjecture. For illustration, we present the following example. Let $X$ be a three-fold obtained by intersecting two generic linear hypersurfaces in a product of five $\IP^1$ spaces. This corresponds to the configuration matrix with identifier 7447 in the CICY list,
\begin{equation}
\label{conf}
X~=~\cicy{\IP^{1}\\[0pt]\IP^{1}\\[0pt]\IP^{1}\\[0pt]\IP^{1}\\[0pt]\IP^{1}}{1&1\\[0pt]1&1\\[0pt]1&1\\[0pt]1&1\\[0pt]1&1}^{5,45}~
\end{equation}
The manifold admits flops in all five directions and the flopped spaces are isomorphic to $X$. This can be seen by explicitly constructing the flops. Let $x=[x_0:x_1]$ denote the coordinates on the first $\IP^1$ and let $y,z,t,s$ be the coordinates of the remaining four $\IP^1$'s. In this notation, the two equations can be written as
\begin{equation}\label{eq}
\begin{aligned}
x_0 P_0(y,z,t,s) &+ x_1 P_1(y,z,t,s) =0~,\\
x_0 Q_0(y,z,t,s) &+ x_1 Q_1(y,z,t,s) =0~.
\end{aligned}
\end{equation}
Since $(x_0,x_1)=(0,0)$ does not correspond to a point in~$\IP^1$, we must have $P_0 Q_1-P_1Q_0=0$. When this condition is satisfied, the equations determine a unique point in the first $\IP^1$, except when $P_0 = P_1=Q_0=Q_1=0$, which happens at $24$ points, when an entire $\IP^1$ is obtained as a solution. Collapsing these $24$ rational curves produces a singular tetra-quadric three-fold $\{P_0 Q_1-P_1Q_0=0\}\subset (\IP^1)^{\times 4}$. The flopped manifold can be constructed by analogy with the Atiyah flop,
\begin{equation}\label{eq_flop}
\begin{aligned}
x_0 P_0(y,z,t,s) &+ x_1 Q_0(y,z,t,s) =0~,\\
x_0 P_1(y,z,t,s) &+ x_1 Q_1(y,z,t,s) =0~
\end{aligned}
\end{equation}
and belongs to the same configuration class \eqref{conf}.  The same discussion can be carried out in all other four directions, leading to a total of five flops for $X$. Each of the flopped manifolds being isomorphic to $X$, admits itself five different flops, one of which takes it back to $X$. The process can be iterated indefinitely, leading to an infinite number of flops and a fractal-like structure of the extended K\"ahler cone, which consists of the union of the K\"ahler cones of all birational models connected to $X$ through a sequence of flops. A schematic representation of the extended K\"ahler cone is shown in Figure~\ref{KX_7447}, which should be interpreted as a cross-section.

\section{String Theory Applications}
\subsection{Heterotic model building.}
Cohomology computations represent the main limiting factor in the analysis of heterotic string compactifications on Calabi-Yau threefolds with holomorphic vector bundles. 
The new cohomology formulae essentially remove this obstacle, allowing for a quick check of the entire low-energy spectrum. This represents a significant improvement from the standard situation where spectrum checks are usually limited to making sure that the number of chiral families, computed as a topological index, is equal to $3$. 

More concretely, heterotic compactifications with sums of line bundles have proven to be a rich arena for model building leading to a large number of standard-like models with many phenomenologically attractive features \cite{Anderson:2011ns, Anderson:2013xka, Constantin:2018xkj}. In this context cohomology constraints can be imposed to ensure: three families of quarks and leptons, the presence of a Higgs field, no exotic matter charged under the SM gauge group, a hierarchy of Yukawa couplings compatible with the observations, the absence of proton decay inducing operators, the existence of right-handed neutrinos, Yukawa unification etc. 

While the cohomology formulae could, in principle, be inverted to infer in a bottom-up fashion the topological data of the compactification directly from the low-energy physical requirements, this approach is not feasible in practice. The constraints take the form of a system of linear, quadratic and cubic diophantine equations which are hard to solve explicitly or even undecidable~\cite{Matijasevic:1970aaa,Halverson:2018cio,Halverson:2019vmd}. However, the cohomology formulae can be used in conjunction with heuristic methods of search, such as Reinforcement Learning and Genetic Algorithms, to identify models resembling the Standard Model at a detailed level of analysis. These methods have recently been shown to be extremely efficient in identifying phenomenologically rich patches of the string landscape without systematically scanning over the entire range of solutions \cite{Halverson:2019tkf, Cole:2019enn, Larfors:2020ugo, Constantin:2021for, Krippendorf:2021uxu, Abel:2021rrj, Abel:2021ddu, Cole:2021nnt}.    

Going beyond abelian bundles, the cohomology formulae discussed above can also be useful in the context of heterotic model building with monad bundles. The monad sequence involves sums of line bundles and their cohomologies feed into the cohomology computations for the monad bundle. 
It is conceivable that simple analytic formulae for bundle cohomology exist also in the case of monad bundles, though more work is needed in this direction. However, even at this stage certain necessary conditions required for the consistency of heterotic compactifications with monad bundles, such as bundle stability, can be expressed in terms of the vanishing of line bundle cohomology groups. Such checks are now immediate and have already been used in Refs.~\cite{Constantin:2021for, Abel:2021rrj, Abel:2021ddu}.

\subsection{Moduli stabilisation.}

Most of the cohomology formulae studied so far correspond to manifolds defined at generic points in the complex structure moduli space and indicate that line bundle cohomology is related in a simple way to the topological and algebraic properties of the base manifold. As the complex structure is varied, cohomology dimensions can jump. The example presented in Section~\ref{sec:CS} indicates that such changes are not isolated and are correlated with something drastic happening to the structure of the cohomology cones. Understanding the manner in which cohomology chambers change with complex structure can be a valuable asset for moduli stabilisation. 

The basic idea is that, for certain bundles and certain choices for the complex structure of the base manifold, the latter cannot be altered without ruining the holomorphicity of the bundle, which introduces a potential for the moduli \cite{Anderson:2010mh, Anderson:2011cza}. The presence of such configurations is signalled by certain cohomology jumps. Working out all the loci in complex structure moduli space where a given bundle stabilises some of the moduli is a hard problem, requiring Gr\"obner base calculations that have double-exponential running time~\cite{Moller:1984aa,Halverson:2018cio} and hence are often completely unfeasible. 
The complex structure dependence of the new cohomology formulae can lead to a quick identification of jumping loci and this information can be used in the construction of hidden sector bundles that can stabilise a maximal number of complex structure moduli in this way.

\vspace{12pt}
{\bfseries Acknowledgements.} The work of CB and JG is supported by the NSF grant PHY-2014086. The work of AC is supported by the EPSRC grant EP/T016280/1. The work of FR is supported by startup funding from Northeastern University. We are grateful to John Christian Ottem for useful comments on an earlier version of this article. 



\end{document}